\documentclass[%
 reprint,
 amsmath,amssymb,
 aps,
]{revtex4-1}
\usepackage{graphicx}
\usepackage{dcolumn}
\usepackage{bm}
\usepackage{hyperref}
\newcommand{\pd}{\partial}
\begin{document}

\title{St\"uckelberg Fields on the Effective $p$-brane}

\author{Patrick Cooper}
 \email{pjc370@nyu.edu}
\affiliation{Center for Cosmology and Particle Physics, Department of Physics, New York University, New York, NY, 10003, USA}

\date{\today}

\begin{abstract}
We   demonstrate the one-to-one correspondence between reparametrization invariant effective actions for relativistic  $p$-branes
in flat target space and effective actions for transverse brane perturbations with non-linearly realized Poincar\'e symmetry.
Starting with an action with non-linearly realized symmetry we construct the corresponding reparametrization invariant action by introducing
St\"uckelberg fields. They combine with the transverse modes to form a Lorentz vector. The manifest Lorentz symmetry of the reparametrization invariant action
follows directly from the non-linearly realized Lorentz symmetry of the initial action in terms of the physical modes.
\end{abstract}

\maketitle


\section{\label{sec:level1}Introduction}

Confinement in quantum chromodynamics (QCD) arises due to the formation of
gluonic flux tubes as is nicely visualized by lattice simulations (see, e.g., \cite{Bissey:2006bz}).
The dynamics of the transverse oscillations of a flux tube are
described by the 2D effective field theory of a string. The leading term, known as the
Nambu-Goto action, is simply the area of the dynamical, effectively two-dimensional surface
of the tube, known as the worldsheet. In the QCD case, where the effective string has a finite
width associated to it, there are higher order terms given by combinations of the first
and second fundamental form associated with the induced metric of the worldsheet in an
ambient spacetime,
\begin{eqnarray}
\label{diffaction}
    S_{string} = - \int d^2 \sigma \sqrt{- h} \left(
    l_s^{-2} + \frac{1}{\alpha_0} (K^i_{ab})^2 + \cdots \right).
\end{eqnarray}
Here the worldsheet metric $h_{ab}$, extrinsic curvature $K^i_{ab}$ and all higher order terms are expressed as functions of the embedding coordinates $X^\mu$,
\[
h_{ab}=\partial_a X^\mu\partial_bX_\mu\;,
\]
where the bulk metric is chosen to be $g_{\mu \nu} = \mathrm{diag}(-,+,\cdots,+)$.
This reparametrization invariant worldsheet description is helpful in that we know all
of the local geometric invariants of embedded 2D surfaces, thus we know the most general
local action compatible with the Poincar\'e symmetry of the theory, as
well as diffeomorphism invariance of the worldsheet. On the other hand, as with any gauge symmetry, the
diffeomorphism invariance of the theory leaves us with a huge redundancy in our description and obscures the counting of physical degrees of freedom.
The natural language to describe the string dynamics directly in terms of propagating degrees freedom  is that of Goldstone bosons
(see, e.g., \cite{Dubovsky:2012sh} or \cite{Aharony:2013ipa} for a recent discussion of effective strings from this viewpoint). All of our discussion applies equally well for any $p$-brane, rather than just a string,
so we consider this more general case in what follows.
A  straight $p$-brane spontaneously breaks the target space Poincar\'e group $ISO(1,D-1)$ down to the direct product of Poincar\'e
transformations along the $p$-brane and rotations in the transverse hyperplane,
$ISO(1,p)\times SO(D-p-1)$. The Goldstone Lagrangian can then be written
as a derivative expansion of the form
\begin{eqnarray}
\label{action}
S =  \int d^{p + 1} \sigma (c_1 \partial_a X^i \partial^a X_i +
    c_2 (\partial_a X^i \partial^a X_i)^2 +  \\
    c_3 (\partial_a X^i \partial_b X_i)(\partial^a X^j \partial^b X_j) + \cdots )\; , \nonumber
\end{eqnarray}
where $X^i$ are the dynamical degrees of freedom  corresponding to the $(D-p-1)$ transverse oscillations
of the brane. Both transverse translations and off-diagonal generators of the target space
Lorentz group are realized non-linearly. Non-linearly realized translations imply the shift invariance of the action (\ref{action}).
 Non-linearly realized rotations/boosts in the
the $(aj)$ plane, where $a$ labels a hypersurface tangent to the $p$-brane and $j$ is normal to
this surface in the bulk, act as
\begin{eqnarray}
\label{NLboost}
\delta_{NL} X^i = - \epsilon_{a j} (\delta^{ij} \sigma^a + X^j \partial^a X^{i}) \equiv -\epsilon_{aj}\delta^{aj}_{NL} X^i \; ,
\end{eqnarray}
This transformation law implies an infinite number of relations between
coefficients $c_i$ in front of the individual terms in the action (\ref{action}). This form of the transformation can be deduced by noticing that actions
of the form (\ref{action}) can be obtained by fixing to the static gauge $X^a=\sigma^a$ in the reparametrization invariant action
(\ref{diffaction}). Then the transformation (\ref{NLboost}) is a combination of a boost and a compensating
diffeomorphism, required to satisfy the gauge condition.

In principle, one may consider actions of the form (\ref{action}) invariant under (\ref{NLboost}) on its own  without any reference to  gauge fixing.
It is natural to expect, however, that all of them can be obtained by gauge fixing some reparametrization invariant action
invariant under the linearly realized Poincar\'e group.

This expectation was challenged recently in Ref.~\cite{Gliozzi:2012cx}, where an inductive procedure was developed
to construct actions of the form (\ref{action}) invariant under (\ref{NLboost}) starting with an arbitrary monomial ``seed" term with a minimal number of $X^i$'s, invariant under
$\delta^{aj}X^i= \delta^{ij}  \sigma^a $. It is convenient to use the ``scaling" ($\pd^n X^m \implies n-m$)
of an operator to label operators that do not mix under (\ref{NLboost}).
Initially, it was claimed that this way, already at scaling two, one may construct actions invariant under   (\ref{NLboost}), which do not correspond
to any local geometric invariant. This claim was later revoked in \cite{Meineri:2013ew}, however the general question remains.

The purpose of this note is to show that the natural expectation is correct and there is a one-to-one correspondence between $p$-brane actions
of the form (\ref{diffaction}) and (\ref{action}). To achieve this we use the St\"uckelberg technique
to (re)introduce reparametrization invariance. We find that in the presence of the symmetry defined by
(\ref{NLboost}) St\"uckelberg fields automatically provide the proper degrees of freedom to restore the
manifest $D$-dimensional Poincar\'e invariance. That is, the non-linear invariance of the initial
Lagrangian translates into a linear Poincar\'e symmetry of the St\"uckelberg action.
We follow the variation of the St\"uckelberg procedure described in
\cite{Dubovsky:2004sg}, which is most convenient for our purposes.
Namely, to introduce  diffeomorphism invariance in any theory, one
replaces all the fields in the action with their image under a diffeomorphism
\begin{eqnarray}
    \sigma^{a} \rightarrow \eta^{a}(\vec{\sigma}) \; , \nonumber
\end{eqnarray}
and adds $\eta^a$ to the set of dynamical fields. The resulting action is equivalent to the initial one and invariant under
coordinate transformations, $\sigma^a \rightarrow \sigma'^a(\vec{\sigma})$, provided the new fields transform as
\begin{eqnarray}
    \eta^{a} \rightarrow \left( \sigma' \circ \eta \right)^{a} \; , \nonumber
\end{eqnarray}
with $\circ$ denoting the composition of the diffeomorphism $\sigma'$ with $\eta$.
The St\"uckelberg fields $\eta^a$  do not transform as scalars with respect to coordinate transformations,
however  the inverse components of the diffeomorphism induced by $\eta^a$ do.  That is,
if we perform a field redefinition from $\eta^a$ to $\xi^{a}$ such that
\begin{equation}
\label{etadef}
\xi^{a}(\vec{\eta}(\vec{\sigma})) = \sigma^{a}
\end{equation}
 then
\begin{eqnarray}
    \xi^{a}(\vec{\sigma}) \rightarrow \left( \xi \circ \sigma' \right)^{a} = \xi^a(\vec{\sigma}'(\vec{\sigma})) \nonumber \; ,
\end{eqnarray}
which is the transformation rule for a scalar.
As we show, these fields, when packaged with the physical transverse oscillations
of the worldsheet, form a $D$-dimensional Lorentz vector,
$X^{\mu}(\vec{\sigma}) \equiv (\xi^a(\vec{\sigma}),X^i(\vec{\sigma}))$.
This proves the one-to-one correspondence between actions (\ref{diffaction}) and (\ref{action}).

\section{The General Proof}

We start with  the action (\ref{action}), depending
only on the physical transverse degrees of freedom of the worldsheet (``Goldstone fields").
As explained above, it should be invariant under a non-linear Lorentz transformation, i.e.
\begin{eqnarray}
    \delta^{a j}_{NL} S \left[ \vec{X}(\vec{\sigma}) \right]
    =
    \int d^{p+1} \sigma  \frac{\delta S[ \vec{X}(\vec{\sigma}) ]}{\delta X^i}
    \delta^{a j}_{NL} X^i(\vec{\sigma}) \nonumber = 0 \; , \nonumber
\end{eqnarray}
where $\delta^{aj}_{NL} X^i(\vec{\sigma})$ is given by equation (\ref{NLboost}).
Our goal is to check that this symmetry translates into an invariance under linear Lorentz transformations after the
reparametrization invariance is (re)introduced via the St\"uckelberg technique.
The  St\"uckelberg prescription is to replace the action with a new
one defined as
\begin{eqnarray}
    S[\vec{X}(\vec{\sigma})] \rightarrow S[\vec{X}(\vec{\eta}(\vec{\sigma}))] \; . \nonumber
\end{eqnarray}
This new action is a functional depending on $(D-p-1)$ Goldstones fields $X^i$ and $(p+1)$ St\"uckelberg fields $\eta^a$.
Equivalently, we can make a field redefinition and treat  this action as a functional depending on  the Goldstone fields $X^i$ and
the inverse St\"uckelberg fields, $\xi^a(\vec{\sigma})$, defined above. By construction, this functional is reparametrization invariant with
both $X^i$ and $\xi^a(\vec{\sigma})$ transforming as scalars.
 Our claim is that in addition, as a consequence of the non-linear Lorentz symmetry (\ref{NLboost}), this action is also invariant under a {\it linearly} realized
 Lorentz symmetry with $X^i$ and $\xi^a$ transforming as components of a Lorentz vector $X^\mu=(\xi^a, X^i)$,
\begin{eqnarray}
\label{linearvar}
    \delta_{L} X^{\mu} = -\epsilon_{a j} (-\xi^a g^{j \mu} + X^j g^{a \mu} ) \equiv -\epsilon_{a j} \delta^{a j}_L X^{\mu} \; .
\end{eqnarray}
To prove this, lets us  show  that
$\delta_{NL} S[X(\vec{\eta}(\vec{\sigma}))] = 0$ implies $\delta_{L} S[\vec{X}(\vec{\sigma}),\vec{\xi}(\vec{\sigma})] = 0$.
The latter variation is equal to
\begin{gather}
\label{var}
-\epsilon_{a j} {\delta S[\vec{X}(\eta(\vec{\sigma}))]\over \delta X^i}
\left(\delta^{a j}_L X^i(\vec{\eta}(\vec{\sigma})) +
    \frac{\partial X^i (\vec{\eta}(\vec{\sigma}))}{\partial \eta^b(\vec{\sigma})}
    \delta^{a j}_L \eta^b(\vec{\sigma})\right) \; .
\end{gather}
To evaluate $\delta^{a j}_L \eta^b(\vec{\sigma})$ let us take the variation of (\ref{etadef}), which gives
\begin{eqnarray}
\label{etavar}
    - \frac{\partial \eta^b (\vec{\sigma})}{\partial \sigma^c} \delta \xi^c (\vec{\eta}(\vec{\sigma}))
    = \delta \eta^b(\vec{\sigma}) \; .
\end{eqnarray}
After plugging (\ref{etavar}) and (\ref{linearvar}) into the variation (\ref{var}) and making use of the chain rule, the variation
(\ref{var}) takes the form
\begin{eqnarray}
   \epsilon_{a j} {\delta S[\vec{X}(\eta(\vec{\sigma}))]\over \delta X^i} \left( \delta^{i j} \sigma^a + X^j(\vec{\eta}(\vec{\sigma})) \,
    \frac{\partial X^i(\vec{\eta}(\vec{\sigma}))}{\partial \sigma_a} \right) \; ,
    \nonumber
\end{eqnarray}
which is simply $ \epsilon_{a j} \delta^{a j}_{NL} S \left[ \vec{X}(\vec{\eta}(\vec{\sigma}) )\right]$ and, as a result, vanishes as a consequence of
the non-linearly realized Lorentz symmetry of the original action.
Thus $\delta^{a j}_L$ is equivalent to $\delta^{a j}_{NL}$ when the linear transformation
is seen as acting on the appropriate combination of Goldstone and inverse
St\"uckelberg fields.
This completes the proof that a \emph{generic} effective Lagrangian invariant under non-linear Lorentz symmetry can
be obtained as a result of gauge fixing to the static gauge the corresponding reparametrization invariant Lagrangian
with linear Lorentz symmetry. After the St\"uckelberg procedure, the non-linear symmetry of the original Lagrangian
translates directly into in the linear Poincar\'e invariance, transforming fields $X^{\mu}$ as a vector.
Hence, as was natural to expect, by gauge fixing generic geometric actions of the form (\ref{diffaction}) one obtains an exhaustive list
of actions invariant under the non-linearly realized  Lorentz (\ref{NLboost}) and shift symmetry.

\section{A Concrete Example}
Given the somewhat abstract nature of the general proof of the previous section, we feel it is instructve to follow in more details
 how the St\"uckelberg procedure works in a concrete example. For simplicity, let us consider the case when the action has scaling zero,
 {\it i.e.} the Lagrangian is a function of the first derivatives $\partial_c X^i$ only. The first step of the St\"uckelberg procedure
 results in an action of the form
\begin{eqnarray}
    S[\vec{X}(\vec{\eta}(\vec{\sigma}))] = \int d^{p+1} \sigma \mathcal{L}
     (\partial_c X^i (\vec{\eta}(\vec{\sigma}))) \; , \nonumber
\end{eqnarray}
where partial derivatives $\partial_c$ will always refer to differentiation with respect to the variable of integration
unless stated otherwise.  To introduce the inverse St\"uckelberg fields, let us change the integration variable,
 $\sigma^a \rightarrow \eta^a(\vec{\sigma}) \equiv \alpha^a$, so that
$\sigma^a = \xi^a(\vec{\alpha})$. To make sense of the argument of the Lagrangian in these variables, we introduce the inverse
Jacobian of this transformation
\begin{equation}
    J(\vec{\alpha})^c_b \equiv [(\partial \xi)^{-1}]^c_b \; ,
\end{equation}
where $-1$ is understood as inverting the matrix of first partials of the variables $\vec{\xi}$.
This substitution leaves us with
\begin{eqnarray}
    S[\vec{X},\vec{\xi}]  = \int d^{p+1} \alpha\;
   \mathrm{det}\left({\partial \xi} \right)
   \mathcal{L}( J(\vec{\alpha})^c_b \partial_c X^i(\vec{\alpha}) )
     \; . \nonumber
\end{eqnarray}
Let us check that this Lagrangian is invariant under linear Lorentz
transformations on the vector $X^{\mu}(\vec{\alpha}) \equiv (\xi^a(\vec{\alpha}), X^i(\vec{\alpha}))$,
provided the original  action is invariant under non-linear Lorentz transformations. Under  a rotation
in the $(aj)$ plane (omitting for brevity the parameter $\epsilon_{a j}$) the action transforms as
\begin{gather}
\label{Lvariation}
    \delta^{a j}_{L}S[\vec{X},\vec{\xi}] =
    \int d^{p+1} \alpha \frac{\partial L(\partial_c \xi^b, \partial_c X^i)}{\partial(\partial_f X^k)}
    \delta^{a j}_{L} (\partial_f X^k)  \\
    + \frac{\partial L(\partial_c \xi^b, \partial_c X^i)}{\partial(\partial_f \xi^h)}
    \delta^{a j}_{L} (\partial_f \xi^h) \nonumber\;,
\end{gather}
where
\begin{eqnarray}
    L(\partial_c \xi^b, \partial_c X^i) = \mathrm{\det}( \partial_e \xi^d )
    \, \mathcal{L}( J(\vec{\alpha})^c_b \partial_c X^i(\vec{\alpha}) )\;.  \nonumber
\end{eqnarray}
At this point it will be useful to introduce some notation to clean up our calculation.  Let us define
\begin{equation}
    \frac{\partial \mathcal{L}(J^c_b \partial_c X^i)}{\partial (J^{c'}_g \partial_{c'} X^l)} \equiv \mathcal{L}_{(g l)} \; . \nonumber
\end{equation}
Now we notice the following chain rules, making use of this notation to change the differential operators acting
on the Lagrangian to be with respect to its argument
\begin{gather}
    \frac{\partial}{\partial(\partial_f X^k)} \rightarrow
    \frac{\partial(J^c_g \partial_c X^l)}{\partial (\partial_f X^k)}
    \mathcal{L}_{(g l)} = \nonumber \\
    = J^c_g \delta^f_c \delta^l_k \mathcal{L}_{(g l)}= J^f_g \mathcal{L}_{(g k)} \; ,
    \label{chainrulex}
\end{gather}
as well as
\begin{gather}
\label{chainrulexi}
    \frac{\partial}{\partial (\partial_f \xi^h)} \rightarrow
    \frac{\partial (J^c_g \partial_c X^l)}{\partial (\partial_f \xi^h)} \mathcal{L}_{(g l)}
    = \partial_c X^l (-J^c_h J^f_g)  \mathcal{L}_{(g l)} \; .
\end{gather}
The integrand of the first term in (\ref{Lvariation}), using equation (\ref{linearvar}) for
the linear variation of $X^k$ as well as the transformation (\ref{chainrulex}), becomes
\begin{eqnarray}
    -\mathrm{det}(\partial \xi)
    \frac{\partial \mathcal{L} }{\partial \left(  \partial_f X^k \right)} \partial_f \xi^a \delta^{j k}
    =
    -\mathrm{det} ( \partial \xi ) \mathcal{L}_{(a k)} \delta^{j k}
\label{term1}
\end{eqnarray}
The $\delta (\partial\xi)$ term in (\ref{Lvariation}) breaks off further into two terms since both the determinant and the original
Lagrangian depend on $\partial \xi$. The differentiation of the determinant along with the variation of the inverse
St\"uckelberg field yields:
\begin{equation}
\label{term2}
    \mathrm{det} ( \partial \xi ) J^f_h \partial_f X^j \mathcal{L} \; g^{a h} \; .
\end{equation}
The differentiation of the original Lagrangian, after implementing (\ref{chainrulexi}), leaves us with
\begin{eqnarray}
\label{term3}
    - \mathrm{det} ( \partial \xi ) ( J^f_g \partial_f X^j )
    ( J^c_h \partial_{c} X^l ) \mathcal{L}_{(g l)} g^{a h} \; .
\end{eqnarray}
The variation of the action is now the sum of equations (\ref{term1}), (\ref{term2}) and (\ref{term3}).  We perform one more change
of variables to get this into a form that we can juxtapose against the gauge fixed action in order to
make use of the fact that the original Lagrangian possessed non-linear symmetry.  Since
every term has $d^{p+1} \alpha\;\mathrm{det}\left( \partial \xi/\partial \alpha\right)$ we simply make $\xi$
our variable of integration. This also implies $J^a_b \partial_a X^i \rightarrow \partial_b X^i$ when we change notation $\partial_a \rightarrow \partial/\partial \xi^a$.
The variation becomes:
\begin{equation}
\label{diffeq}
    \int d^{p+1} \xi \left(
    -\mathcal{L}_{(a k)} \delta^{j k}
    + \partial_h X^j \mathcal{L} g^{a h}
    - \partial_g X^j \partial_h X^l \mathcal{L}_{(g l)} g^{a h} \right) \; .
\end{equation}
The second term we integrate by parts:
\begin{gather}
    \partial^a X^j \mathcal{L} \rightarrow
    - X^j \partial^a \mathcal{L} = - X^j \partial^a \partial_d X^k \mathcal{L}_{(d k)} \; . \nonumber
\end{gather}
Now we rename the indices $d$ and $g$ of the second and third term respectively both to $h$. We also
rename the index $l$ of the last term $k$ and use a Kronecker delta to rename the index
$a$ of the first term $h$. We can factor out the common derivative of
the Lagrangian leaving us with
\begin{widetext}
    \begin{eqnarray}
        -\int d^{p+1} \xi \mathcal{L}_{(h k)}
        \left( \delta^a_h \delta^{j k} + X^j \partial^a \partial_h X^k
        + \partial^a X^j \partial_h X^k \right) \; . \nonumber
    \end{eqnarray}
\end{widetext}
which by equation (\ref{NLboost}) is proportional to
\begin{equation}
    \int d^{p+1} \xi \mathcal{L}_{(h k)} \; \delta^{a j}_{NL} (\partial_h X^k)
    = \delta^{a j}_{NL} S[\vec{X}] \; .
\end{equation}
This vanishes by our initial assumption. Thus any Lagrangian with this non-linear symmetry can be turned into a
manifestly Poincar\'{e} and diffeomorphism invariant action. One simply restores diffeomorphism invariance
with St\"uckelberg fields and identifies the appropriate degrees of freedom which combine with the physical
degrees of freedom to form the embedding coordinates of the worldsheet in spacetime. The original
Lagrangian can always be interpreted as the static gauge of this procedure.

As an aside, we emphasize that non-linearly realized Lorentz symmetry implies, restoring the variation parameter $\epsilon_{a j}$,
\begin{eqnarray}
    \epsilon_{a j} \int d^{p+1} \xi \left(
    -\mathcal{L}_{(a k)} \delta^{j k}
    + \partial_h X^j \mathcal{L} g^{a h}
    - \partial_g X^j \partial_h X^l \mathcal{L}_{(g l)} g^{a h} \right) \nonumber \\
    = 0 \nonumber
\end{eqnarray}
This leaves us with a differential equation for $\mathcal{L}$. That is, the integrand must be
proportional to a total derivative. Since $\mathcal{L}$ is a function of
$\partial X$ only, the only total derivative compatible with the symmetries of the Goldstone Lagrangian
that the integrand could be proportional to is a suitable contraction of $\partial^a X^j$. By shifting the
Lagrangian by a constant, the second term in the integrand can be tuned to cancel this total derivative so
our differential equation can be defined by setting the integrand above to 0.
To solve this relation, we can write the Lagrangian as
\begin{equation}
    \mathcal{L}(\partial_c X^k) = \sum_{c, k} \sum_n a^{(c, k)}_n (\partial_c X^k)^n \; . \nonumber
\end{equation}
again, noting that the coefficients $a_n^{(c k)}$ aren't completely arbitrary in $c$ and $k$, since the Lagrangian still must
respect the manifest $ISO(1,p) \times SO(D-p-1)$ symmetry. The derivatives and products of $\partial X$ then provide us with
recurrences relations between the powers of $\partial_c X^k$ in the argument of the Lagrangian. The resummation of these terms
is precisely the procedure in \cite{Gliozzi:2012cx} to show that $\mathcal{L}$ is the invariant area of the $p$-brane.
As a trivial example, for the $0$-brane, the left hand side of this differential equation reads
\begin{equation}
    -\frac{\partial \mathcal{L}(\dot{\vec{X}}(t) )}{\partial \dot{X}^k(t) } \delta^{jk} - \dot{X}^j(t) \mathcal{L}(\dot{\vec{X}}(t))
    + \dot{X}^j \dot{X}^l(t) \frac{\partial \mathcal{L}(\dot{\vec{X}}(t) )}{\partial \dot{X}^k(t)} \delta^{kl} \; , \nonumber
\end{equation}
where the dot implies differentiation with respect to $t$.  To solve this differential equation we use spherical coordinates for the
set of vectors $\dot{\vec{X}}$ since we know that the Goldstone Lagrangian possesses an $SO(D-1)$ symmetry.
Thus the multidimensional differential equation becomes a one dimensional equation for $\mathcal{L}(r)$ with $r^2 = \dot{X} \cdot \dot{X}$
\begin{equation}
    -\frac{\partial \mathcal{L}(r)}{\partial r} - r \mathcal{L}(r) + r^2 \frac{\partial \mathcal{L}(r)}{\partial r} = 0 \; . \nonumber
\end{equation}
\\
whose solution is given by the invariant length of the worldline
\begin{equation}
    \mathcal{L}(\dot{\vec{X}}(t)) = -m \sqrt{1 - \dot{X}^i(t) \dot{X}^i(t)} \; . \nonumber
\end{equation}

\section{Acknowledgments}
I'd like to thank Sergei Dubovsky, Victor Gorbenko, Micha Gorelick, Kilian Walsh and Guido D'Amico for their valuable feedback and discussions on this topic.

\bibliography{diffstueck}

\end{document}